%% file: bare_jrnl.tex
\documentclass[journal]{IEEEtran}
\usepackage{cite}
\usepackage{graphicx}
\usepackage{amsmath}
\usepackage{amssymb}
\usepackage[ruled,lined,linesnumbered]{algorithm2e}
\let\oldnl\nl
\newcommand{\nonl}{\renewcommand{\nl}{\let\nl\oldnl}}

\usepackage{xcolor}
\usepackage{optidef}
\newtheorem{definition}{Definition}[section]
\DeclareMathOperator*{\argmax}{arg\,max}

\newcommand{\ie}{\textit{i.e.,}~}
\begin{document}
\bstctlcite{IEEEexample:BSTcontrol}
\title{Service Function Chaining in MEC: A Mean-Field Game and Reinforcement Learning Approach}

\author{
	    Amine~Abouaomar,~\IEEEmembership{Student~Member,~IEEE,}
		Soumaya~Cherkaoui,~\IEEEmembership{Senior~Member,~IEEE}
		Zoubeir~Mlika,~\IEEEmembership{Member,~IEEE,}
		and~Abdellatif~Kobbane,~\IEEEmembership{Senior~Member,~IEEE,}
		\thanks{A. Abouaomar is with INTERLAB, Engineering Faculty Université de Sherbrooke, Sherbrooke (QC), Canada, he is also with ENSIAS, Mohammed V University in Rabat, Morocco.}
		\thanks{S. Cherkaoui and Z. Mlika are with INTERLAB, Engineering Faculty Université de Sherbrooke, Sherbrooke (QC), Canada.}
		\thanks{A. Kobbane is with ENSIAS, Mohammed V University in Rabat, Morocco.}
}

\markboth{}{}

\maketitle

\begin{abstract}
Multi-access edge computing (MEC) and network virtualization technologies are important enablers for fifth-generation (5G) networks to deliver diverse applications and services. Services are often provided as fully connected virtual network functions (VNF)s, through  service function chaining (SFC). However, the problem of allocating SFC resources at the network edge still faces many challenges related to the way VNFs are placed, chained and scheduled. In this paper, to solve these problems, we propose a game theory-based approach with the objective to reduce service latency in the context of SFC at the network edge. The problem of allocating SFC resources can be divided into two subproblems. 1) The VNF placement and routing subproblem, and 2) the VNF scheduling subproblem. For the former subproblem, we formulate it as a mean field game (MFG) in which VNFs are modeled as entities contending over edge resources with the goal of reducing the resource consumption of MEC nodes and reducing latency for users. We propose a   on a reinforcement learning-based technique, where the Ishikawa-Mann learning algorithm (IMLA) is used. For the later subproblem we formulate it as a matching game between the VFNs and an edge resources in order to find the execution order of the VNFs while reducing the latency. To efficiently solve it, we propose a modified version of the many-to-one deferred acceptance algorithm (DAA), called the enhanced multi-step deferred acceptance algorithm (eMSDA). To illustrate the performance of the proposed approaches, we perform extensive simulations. The obtained results show that the proposed approaches outperform the benchmarks other state-of-the-art methods.
\end{abstract}


\IEEEpeerreviewmaketitle

\section{Introduction}
The fifth generation (5G) of mobile systems is being promoted to accelerate the development of smart cities, not only through improved data throughput but also through support for the expected large amount of connected devices \cite{8985528}. Many of the use cases involving latency-sensitive applications and highly responsive services, are now achievable with less effort and cost \cite{9055745}. To enable such services, 5G relies on emerging technologies such as cloud computing, multi-access edge computing (MEC), and virtualization technologies that can meet the demands of network flexibility and elasticity. MEC on one hand \cite{filali2020multi} emerged to address the limitations of cloud computing primarily in terms of latency. As an extension of the cloud computing paradigm to the edge of the network, MEC enables data processing close to where it is generated \cite{filali2020multi}. In addition, MEC can assist 5G in enabling ultra-reliable and low-latency communication (URLLC), enhanced mobile broadband (eMBB), and support massive machine-like communication (mMTC) \cite{8985528}. On the other hand, virtualization technologies have appeared as a remarkable concept to provide efficient provisioning for 5G and beyond networks through software-defined networking (SDN) and network functions virtualization (NFV) \cite{nour2020computeless, 9263348}.
Using virtualization technologies, physical network components and hardware devices can be abstracted into software called virtual network functions (VNFs) \cite{7534741}. VNFs can be instantiated and executed in the data plane as virtual machines or containers accommodated in devoted infrastructures such as cloud or MEC platforms. Services are often offered as a chain of multiple VNFs to compose a service function chain (SFC).  Nevertheless, resource provisioning of SFCs remains a challenging problem, especially to meet latency and resource consumption requirements. Resource provisioning in the context of SFCs requires addressing multiple subproblems including (i) the VNFs placement, (ii) the VNFs chaining and (iii) the VNFs scheduling \cite{7534741, filali2020multi}. To achieve all the required performance, it is necessary to treat these three subproblems as an indivisible part of SFC resource provisioning process. The SFC resource provisioning problem should be solved regarding different types of resources, namely computation, storage, and transmission. In fact, many works have studied the problem of SFC resource provisioning but considering only computation or transmission, ignoring the storage resources. However, on many occasions, services may require content storage for caching purposes \cite{abouaomar2017caching} or for storing results of processing as in some machine learning paradigms \cite{tak2020federated}. In addition, most existing solutions \cite{pham2017virtual,9060929,9181472,9013429,9075271,9187253,8647858,9062531} assume that the VNFs have equal requirements and are homogeneous within a single MEC node. However, in real use cases, VNFs often require different amounts and types of resources, and have different purposes. In addition, to meet the URLLC requirements of 5G, it is necessary to consider the MEC as an infrastructure to accommodate VNFs.

The VNFs placement problem aims to find a suitable location for the VNFs (VMs/containers) in MEC nodes, where each MEC has physical devices on which VMs/containers can be instantiated according to demand \cite{7534741}. In fact, the advantage of deploying VNFs at the MEC level is the proximity to users, which allows low latency. Proper placement of VNFs is motivated by optimizing resource consumption or overall network performance, while minimizing the cost (in terms of energy consumption, latency, and financial) or penalties for service level agreement violations. Although these parameters are quite appropriate for NFVs, most of the work proposed in this context focuses only on reducing the cost of deployment, improving QoS and availability, while neglecting the constraints related to MEC itself. These constraints are associated with service latency and resource consumption. The SFC chaining subproblem seeks to find the optimal VNFs path to transmit the processed packets when the number of transmission links or their capacities are limited. It is addressed within cases where the VNFs is deployed/duplicated on several network sites. The two subproblems of VNFs placement and chaining are often addressed jointly in the literature \cite{pham2017virtual,9060929,9181472,9013429,9075271,9187253,8647858,9062531} and many approaches have been proposed for this purpose. We can classify these approaches as heuristic approaches \cite{8892907}, game theory-based approaches \cite{pham2017virtual}, machine learning-based approaches \cite{9187253} and linear programming-based approaches \cite{9181472}. However, these works concentrate on cost minimization under service level agreement constraints without considering MEC constraints.
Finally, the SFC scheduling subproblem should be solved considering different MEC nodes, since SFCs are often running within a shared physical network and thus, several VNFs can be hosted in the same MEC node \cite{7534741}. The VNFs scheduling subproblem can be defined as finding the sequence in which the collocated VNFs on each MEC node will be executed in order to achieve the minimum SFC execution time (i.e., the experienced delay between the execution of the source VNF and the completion of the destination VNF). This problem is in the NP-complete complexity class and thus cannot be solved in polynomial time unless $P=NP$ \cite{pham2017virtual}.

\begin{figure*}[!t]
    \centering
    \includegraphics[width=.7\linewidth]{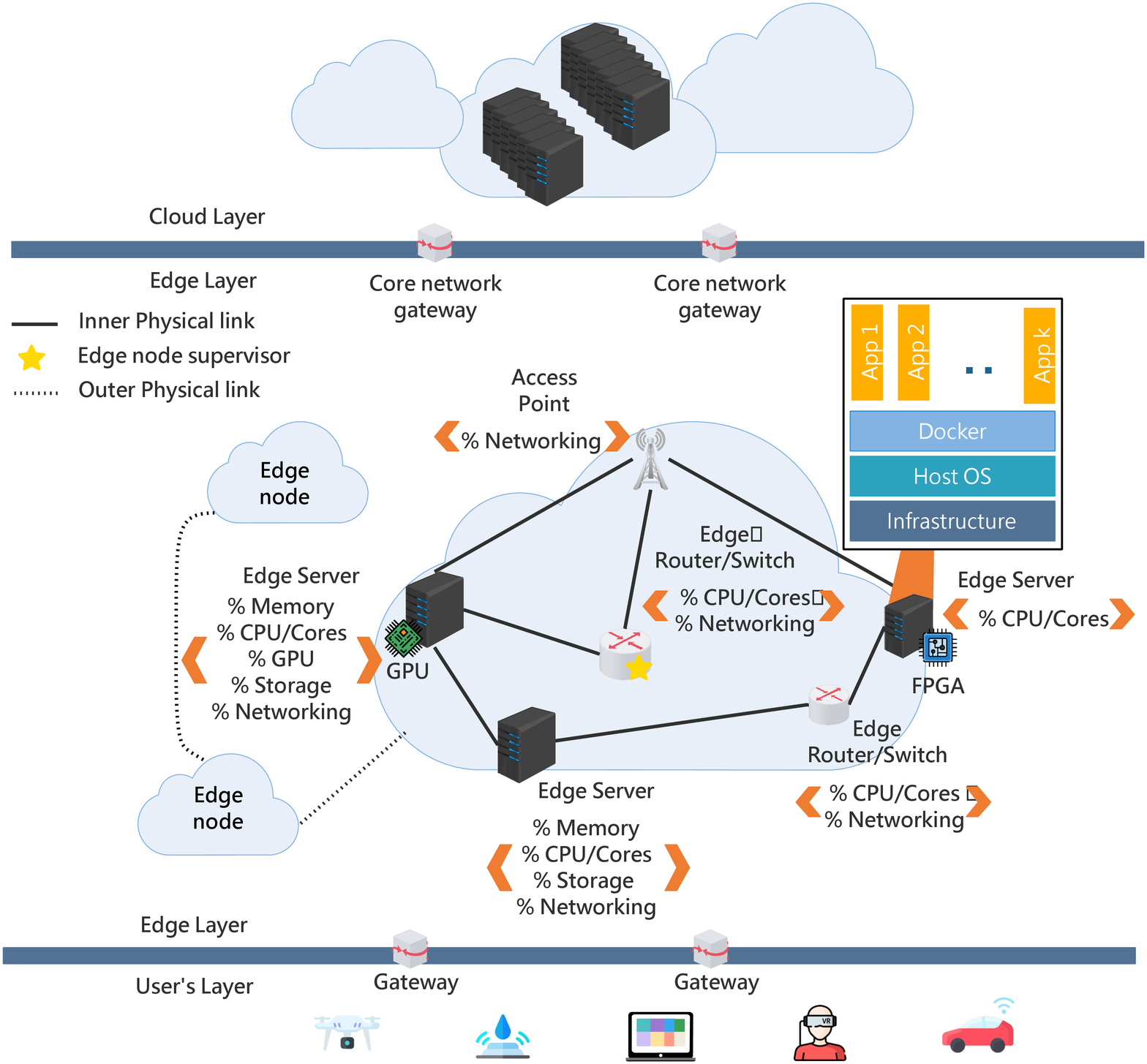}
    \caption{\label{fig:edge-arch} An illustration of the considered MEC architecture and the different system's entities.}
\end{figure*}

To summarize, the main contributions of this paper are synthesized as follows:
\begin{itemize}
    \item We study the SFC resource provisioning problem as an integral part in which we consider the subproblems of VNFs placement, VNFs chaining, and VNFs scheduling with the aim to reduce the overall latency and resource consumption while considering different types of resources, namely computational, storage and transmission resources.
    
    \item We model the VNF placement and chaining subproblems on different MEC nodes as a mean-field game (MFG) to provide the adequate placement and chaining of VNFs on different MEC nodes. The game involves heterogeneous VNFs that require different amounts and types of resources. We also consider a stochastic arrival of requests for an enhanced resource provisioning results.
    
    \item We model the VNF scheduling as a matching game to address the problem of multiple heterogeneous VNFs running within the same MEC node. We extend the classical deferred acceptance algorithm, and we proposed the enhanced multistage deferred acceptance algorithm (eMSDA) to support constraints on processing completion, routing time, and resource consumption.
    
    \item We provide a theoretical study for both games by proving their stability and equilibrium.
\end{itemize}

The remainder of this paper is structured as follows. Section II presents the system model. Subsequently, we formulate the problem of resource provisioning and we present the different system requirements in Section III. In Section IV, we provide the details about the proposed solutions for the VNFs placement, chaining, and scheduling. The performance of the system is simulated in Section V. Last but not least, the related works are discussed in Section VI. Finally, we conclude the paper in Section VII.

\input{system_model}

\input{problem_formulation}

\input{proposed_solution}

\input{simulations}

\input{related_works}

\section{Conclusion}
In this paper, we studied VNF resource provisioning for SFC in a MEC context with the goal of reducing latency. We addressed the full problem of resource provisioning through addressing the VNF placement, the VNF routing and the VNF scheduling. As for the VNF placement and routing problem, we formulated it as a mean field game where VNFs are the players that are competing for MEC resources with the goal of reducing the resource consumption of MEC nodes and decreasing latency for end users. We exploited the Ishikawa-Mann iterated learning algorithm to place and chain the VNFs. We formulated the VNFs scheduling problem as a many-to-one matching to match multiple VFNs to MEC node resources, and we proposed a modified version of the classical deferred acceptance algorithm. We performed extensive simulations to prove the effectiveness of our proposed approaches, which were found to outperform the benchmark approaches in the studied scenarios. In future work, we will investigate the possibility of testing the proposed approach on a real testbed to compare the theoretical results with the experimental results.


\bibliography{references} 
\bibliographystyle{IEEEtran}

\appendices
\section{Proof of BRC second-order derivative.}
\label{app:bcp-deriv}
Here we prove the existence of NE by proving that the second-order derivative of the BRC is negative.
\begin{equation}
    \begin{split}
        \partial_{a_{i}}r_{v}^{*}(a_j, m_{(v,-j)})=\frac{\alpha_i}{v}\left(\frac{\beta a_{\beta-1}^{j}m_{v}^{\beta}-a_{j}^{\beta}\left(\frac{\beta}{v}a_{j}^{\beta-1}\right)}{m_{v}^{2\beta}}\right)-\delta_{(i,j)}\\
        =\beta a_{j}^{\beta-1}\times\frac{\alpha_i}{v}\left(\frac{m_{v}^{\beta}-\frac{a_{j}^{\beta}}{v}}{m_{v}^{2\beta}}\right)-\delta_{(i,j)}
    \end{split}
\end{equation}
from (\ref{eq:mf_term}), we can write:
\begin{equation}
    m_{v}^{\beta}-\frac{a_{j}^{\beta}}{v}=\mathlarger{\sum_{\substack{j'\in\mathcal{F}_i\\j\neq j'}}}a_{j'}^{\beta}
\end{equation}
since $a_{j'}$ is independent of $a_j$, we can develop the second derivative of the payoff function $r_{v}^{*}\left(a_j,m_{(v,-j)}\right)$ as follows:

$$\partial_{a_j}^{2}r_{v}^{*}\left(a_j,m_{(v,-j)}\right)=\\$$
\begin{equation}
\begin{split}
    \frac{\alpha_i}{v}\left(\frac{
        \beta\left(m_{v}^{\beta}-a_{j}^{\beta}\right) \left[\left(\beta-1\right)a_{j}^{\left(\beta-2\right)}m_{v}^{2\beta}-a_{j}^{2\left(\beta-1\right)}m_{v}^{\beta}\left(\frac{2\beta}{v}\right)\right]
    }{m_{v}^{4\beta}}\right) \\
    =\frac{\alpha_i}{v}\left(m_{v}^{\beta}-\frac{a_{j}^{\beta}}{v}\right)\times\frac{a_{j}^{\left(\beta-2\right)}}{m_{v-}^{3\beta}}\times\left(\left(\beta-1\right)m_{v}^{\beta}-\frac{2\beta}{v}a_{j}^{\beta}\right)
\end{split}
\end{equation}
For given values of $\beta\in[0,1]$, the second derivative $\partial_{a_j}^{2}r_{v}^{*}\left(a_j,m_{(v,-j)}\right)$ is negative.
\section{Minimum and Maximum Quota constraints}
\label{app:min-quota}
Let us consider the following scenario with the set of VNFs $F_i=\{f_{1}, f_{2}, f_{3}\}$ and the set of ENs $\mathcal{E}=\{e_1, e_2, e_3\}$. And let us assume that the maximum quota for all the ENs, $q_{i}^{max}=2$ and the minimum quota, $q_{i}^{min}=1$. If we consider that all the ENs share the same preference list $\succ_{f}$, and if we consider that $e_1 \succ_f e_2 \succ_f e_3$ and $f_1 \succ_{f_i} f_2 \succ_{f_i} f_3$, applying the classical DAA gives that (1) $\mu(f1) = \{e_1, e_2\}$; (2) $\mu(f2) = \{e_3\}$; (3) $\mu(f3) = \emptyset$. This breaks the minimum quota rules.

\ifCLASSOPTIONcaptionsoff
  \newpage
\fi

\end{document}

%% file: system_model.tex
\section{System Model}

In this section, we present the considered entities of the system model, followed by the system description, namely, service requests and the physical resources of MEC nodes. Table \ref{table_definition} summarizes the important notations used in the system model.

\input{summary_table}

For sake of clearness, in what follows, we denote by the index $i$ represents the MEC nodes, $j$ represents the users, $k$ represents the services, and $v$ represents the VNFs.

\subsection{Physical network substrate}
In this work, we consider a slotted system with $t \in \mathbb{N}$ that represents the time slots, where $\mathbb{N}$ represents the set of natural numbers.
We consider a MEC network consisting of $N$ edge nodes (ENs) $\mathcal{E}=\{e_1,\ldots,e_N\}$ distributed over a geographical area and interconnected through physical links (PhyL) having transmission capacity of $L_{(i,i')}$ with $(i,i'\in \mathcal{E})$ in $packet/t$ for horizontal data exchanges. Fig.~\ref{fig:edge-arch} illustrates the considered MEC architecture where we consider three layers. The bottom layer represents the users who communicate with the edge layer through edge gateways. Gateways forward the requests to the associated edge node. In this paper, we consider that the network is supervised by an SDN controller. Both user's layer and edge layer belong to the edge of the network and communicate with the top layer that consists of the core network layer. An example of users could be an extended reality equipment \cite{lhazmir2019performance}, unmanned aerial vehicles \cite{9062531}, or connected vehicles \cite{azizian2017vehicle}. We denote the set of users associated to a given EN by $\mathcal{U}_i=\{u_{(i,1)}, \ldots, u_{(i,J)}\}$. Each $i\in\mathcal{E}$ is an aggregation of heterogeneous edge devices such as edge servers, routers, access points, and even eNodeBs/gNodeBs. These devices host a set of VNFs $\mathcal{F}_i=\{f_{(i,1)},\ldots,f_{(i,v)}\}$. Without loss of generality, we consider the placement of the VNFs within a given EN regardless of the specific assignment between the VNFs and the devices of the corresponding EN. Moreover, in this paper, we leverage our previous works \cite{abouaomar2019resources,9326402} on resource representation to allow the EN to exchange their resource availability status and we consider the overall available resources of the EN. Each VNF requires an amount of resource to perform its tasks in terms of computing, storage, and transmission. The available resource at the EN $i$ at time-slot $t$ is given by,
\begin{equation}
    \alpha_i(t)=\langle c_{i}(t), s_{i}(t), \omega_{i}(t)\rangle    
\end{equation}
where $c_{i}(t)$ is the available computational resource, $s_{i}(t)$ the available storage resource, and $\omega_i(t)$ the available transmission resource. Services are represented by the set $\mathcal{S}=\{s_{1},...,s_{O}\}$ in which a single service $s_k$ is given by the following tuple: 
\begin{equation}
    s_{k}=\langle F_{k},f_{src}^{k},f_{dest}^{k},L_{k}\rangle,
\end{equation}

\begin{figure}[!t]
    \centering
	\includegraphics[width=\columnwidth]{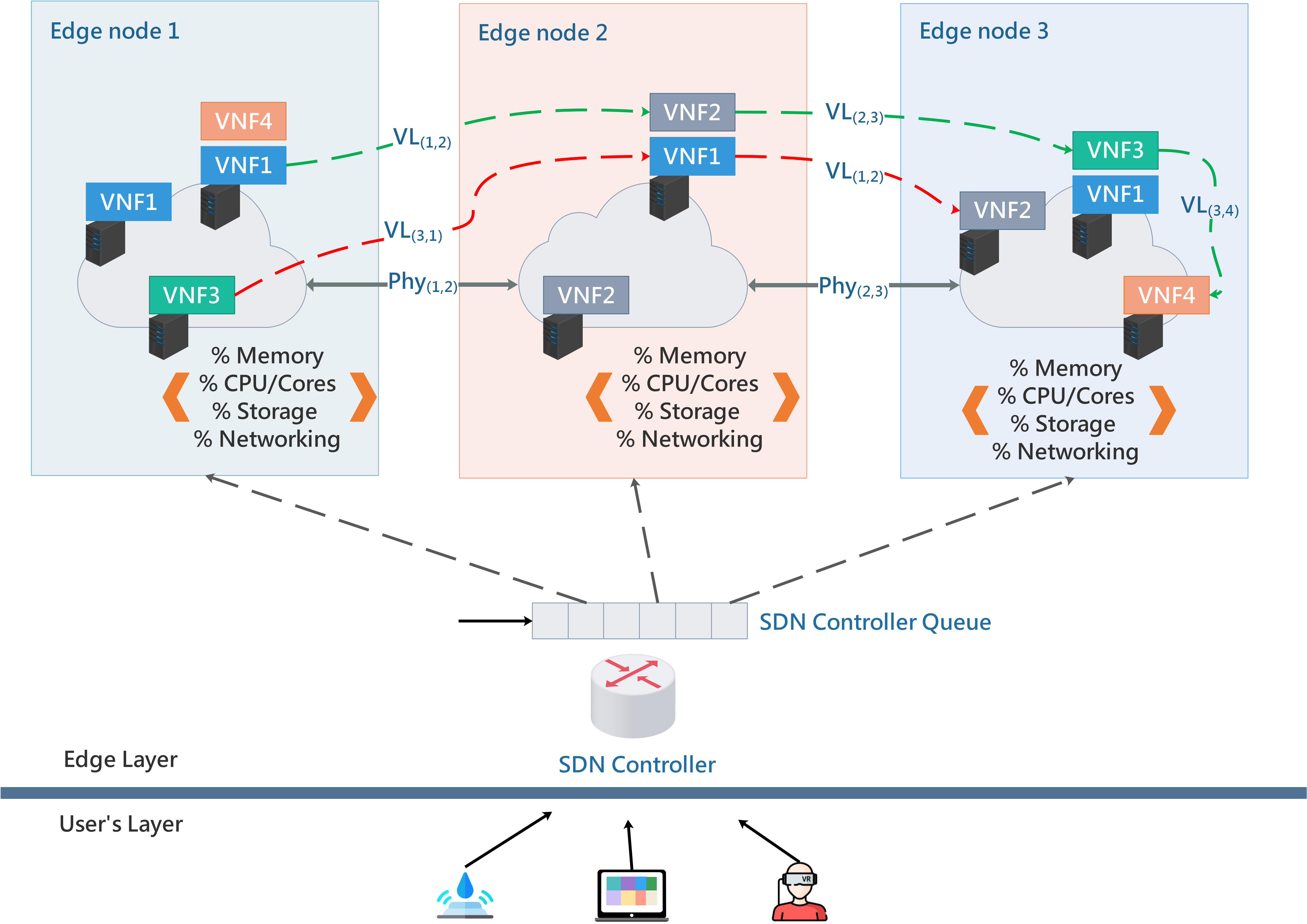}
    \caption{\label{fig:exmpl-sfc} Example of SFC deployment over multiple edge nodes.}
\end{figure}
\noindent where $F_{k}$ denotes the set of VNFs composing the SFC $s_k$ in the ascending order, as depicted in Fig. \ref{fig:exmpl-sfc}. The source and destination VNFs of SFC $s_k$ are given by $f_{src}$ and $f_{dest}$, respectively. The virtual links (VLs) between the VNFs of the SFC $s_k$ is denoted as $L_k$ and is represented with $L_{k}=\{l_{(v,v^{'})} |(v,v^{'}) \in \mathcal{F}_{k}~\text{with}~v \neq v^{'} \}$. We also consider that VNFs have different requirements, 
\begin{equation}
    \bar{\alpha}_v=\left\langle c_{(i,j)}^{v}, s_{(i,j)}^v, \omega_{(i,j)}^{v}\right\rangle,
\end{equation}
\noindent where, $c_{(i,j)}^{v}$ represents the computational requirement, $s_{(i,j)}^v$ the storage requirement, and $\omega_{(i,j)}^{v}$ transmission. We assume that users are already associated to the ENs as proposed in \cite{abouaomar2018matching}. We consider a binary parameter to represent the user-EN assignment $x_{j}^{i}$, defined as:
\begin{subequations}
    \begin{equation}
        x_{j}^{i}=\begin{cases}
            1, & \text{ if user $j$ is associated to EN $i$}, \\
            0, & \text{otherwise}.
        \end{cases}
    \end{equation}
    where
    \begin{equation}
        \mathlarger{\sum_{e_i\in\mathcal{E}}}x_{j}^{i}=1, \forall j\in \mathcal{U}_i.
    \end{equation}    
\end{subequations}
We define the VNF-EN association with the variable $x_{(i,v)}^{f}$ as,
\begin{equation}
    x_{(i,v)}^{f}=\begin{cases}
        1, \text{ if the replica of VNF $v$ is selected in EN $i$}, \\
        0, \text{ otherwise}.
    \end{cases}
\end{equation}

\subsection{Service Requests}
In the literature, many request arrival models have been studied and applied to queuing based systems such as the Poison processes \cite{wolff1982poisson}, statistic arrivals based on observing the state of the system to build an arrival model \cite{gopalan2006statistical,marin2001statistical} and stochastic arrival models \cite{gupta2019stochastic,dehghani2017stochastic,thakral2019matching}. The impact of the request arrival model is crucial, especially when investigating an online-related problem such as in \cite{9060929}. In \cite{9060929}, the authors considered that the requests arrive in a sequence but did not consider a precise model. However, the requests arrival model has a crucial role in studying resource allocation in a MEC architecture. The work in \cite{gupta2019stochastic} investigates the problem of stochastic online matching and considered an independent and identical distributed (IID) request arrival model. This model's requests arrival is stochastic-based that has been studied in a variety of online problems such as the $k$-server problem \cite{dehghani2017stochastic} and matching problem \cite{thakral2019matching}. Adopting an IID model for request arrival with a stochastic framework can improve the quality of the results regarding the resource allocation \cite{gupta2019stochastic}.

In this paper, we adopt IID request arrival model as described in \cite{gupta2019stochastic}. We model the request arrival to denote the means of requests received by an EN using the variable $\sigma$ described as follows:
\begin{equation}
    \sigma=\{\sigma_i,i\in\mathcal{E}\}.
\end{equation}

Let $\rho_{(j,k)}$ represents a request for the service $k$ by the user $j$, and is defined as,
    \begin{equation}
        \rho_{(j,k)}=\left\langle s_k, d_j, T_{out}\right\rangle,
    \end{equation}

\noindent where, $s_k$ is the requested service, $d_j$ is the data packet and $T_{out}$ is the timeout; the time at which the users must complete the service.

\subsection{EN Physical Resources}
As mentioned before, the ENs are formed by aggregating different heterogeneous edge devices such as servers, edge routers, small base stations, or even eNodeBs and gNodeBs. In \cite{abouaomar2019resources, 9326402}, we proposed a resource representation model that allows these heterogeneous devices to exchange information about their resource capabilities in terms of type of the operations that can handle. For instance, the edge routers are more into network operations unlike edge servers that are more into data processing and edge learning. In this paper, we will leverage our proposed resource representation model to get the overall capacities of each EN. In other words, we will focus on the position of the VNFs at the EN in general. The ENs communicate horizontally through PhyL that can generally be wired or wireless. 

\subsubsection{Virtual Links Allocation}
The virtual links must allocate a portion of the physical links resources to transmit the packets. We introduce $r_{(v,v')}^{i}$ as binary parameter to denote whether VNFs are located into different EN. In fact, $r_{(v,v')}^{i}$ defines whether a VL allocation on PhyL is necessary or not. We also define the binary parameter $y_{l}^{z}$ to denote VL-PhyL assignment. $r_{(v,v')}^{i}$ and $y_{l}^{z}$ are defined as follows:
\begin{subequations}
    \begin{equation}
        \label{eq:pvl1}
        y_{l}^{z}=\begin{cases}
            1, & \text{if VL $z$ is allocated on PhyL $(i,i')$}, \\
            0, & \text{otherwise},
        \end{cases}
    \end{equation}
    \begin{equation}
        \label{eq:pvl2}
        r_{(v,v')}^{i}=\begin{cases}
            0, & \text{if VNFs are within the same EN}, \\
            1, & \text{otherwise},
        \end{cases}
    \end{equation}
    \begin{equation}
        \label{eq:pvl3}
        \mathlarger{\sum_{z \in L_k}}~\mathlarger{\sum_{\substack{l \in L_{(i,i')}}}}r_{(v,v')}^{i}\times y_{l}^{z}=1 | \text{ $i\neq i'$ and $v\neq v'$,}
    \end{equation}
\end{subequations}
In case VNFs of the same SFC are located within the same EN, Eq. (\ref{eq:pvl3}) ensures that the PhyL-VL allocation operation is not performed.

\subsubsection{Processing Model}
In the considered system model, each VNF process the forwarded packets using the allocated resource from the EN. Given a request $\rho_{(j,k)}$ with a packet size $d_j$, the processing delay of a packet $j$ is given as follows:
\begin{equation}
    \gamma_{(i,j)}^{v}(t)=\frac{d_j}{\alpha_i(t)}
\end{equation}
The transmission delay between two consecutive VNFs in EN $i$ and $i'$ is given as follows, 
\begin{equation}
    \gamma_{j}^{z}(t)=r_{(v,v')}^{i}\frac{d_j}{l_{(i,i')}^z(t)}
\end{equation}
where $l_{(i,i')}^z$ represents the virtual link $z$ capacity, given as follows,
\begin{equation}
	l_{(i,i')}^z = B.\log_2\left(1 + \frac{P_i\left|g_{(i,i')}\right|^2}{\zeta^2 + \sum_{\substack{i''\in\mathcal{E}\\i''\neq i}} {P_{i''}\left|g_{(i'',i')}\right|^2}}\right),
\end{equation}
where $B$ is the bandwidth, $P_i$ the power of transmission of $i$, $g_{(i,i')}$ the channel gain between the ENs, and $\zeta^2$ the noise variance.
The overall delay experienced from requesting service $k$ is given as follows,
\begin{equation}
    \gamma_{j}(t)=\sigma_{i}\sum_{e_i\in\mathcal{E}} \gamma_{(i,j)}^{v}(t)  + \sum_{l\in L_k} \gamma_{j}^{l}(t) + \gamma_{queue}
\end{equation}

\noindent where $\gamma_{queue}$ is the queuing time at the full controller queue. Without loss of generalities, we consider an $M/M/1$ queue, and we adopt the Little's law to cope with queuing delay.

%% file: summary_table.tex
\begin{table}
\renewcommand{\arraystretch}{1.3}
\caption{Summary of important notations}
\label{table_definition}
\centering
\begin{tabular}{c l}
Symbols & Definition\\
\hline
\hline
$\mathcal{E}$ & Set of edge nodes \\
$\mathcal{U}_{i}$ &  Set of users\\
$\mathcal{F}_{i}$ &  Set of VNFs hosted in EN $i$\\
$\mathcal{F}_{k}$ & Set of VNFs composing the SFC $k$ \\
$\mathcal{S}$ &  Set of SFCs\\
$\rho_{(j,k)}$ & Request for service $k$ by yser $j$\\
$\Bar{\alpha}_{v}$ & Requirement of VNF $v$\\
$c_i$ & Computational capacity of EN $i$ \\
$s_i$ &  Storage capacity of EN $i$ \\
$\omega_i$ &  Transmission of EN $i$ \\
$f_{src}$ & First VNF of SFC $k$ \\
$f_{dest}$ & Destination VNF of SFC $k$ \\
$L_{k}$ & Virtual links of SFC $k$ \\
$x_{j}^{i}$ & User-EN assignment \\
$x_{(i,v)}^{f}$ & VNF-EN assignment \\
$y_{z}^{l}$ & VNF-SFC assignment \\
$z_{z,l}^{i}$ & VNF-SFC-EN assignment \\
$m_v$ & Mean field term\\
$r_{v}^{}$ & Payoff function\\
$a_{j}$ & Strategy taken by VNF $j$ \\
$\mathcal{A}$ & Set of strategies \\
$\lambda$ & Learning rate\\
$\succ_{f_{(i,j)}}$ & Preferences of VNFs overs EN\\
$\succ_{i}$ & Preferences of EN over VNFs\\
$\mu(.)$ & Many-to-One mapping\\
$q_{i}^{min}$ & Quota min\\
$q_{i}^{max}$ & Quota max\\
\hline
\hline
\end{tabular}
\end{table}

%% file: problem_formulation.tex
\section{Problem Formulation}

\subsection{The VFN placement and chaining subproblem}
We formulate the problem as a mean-field resource allocation game \cite{caines2015mean,hanif2015mean} in which a number of VNFs in the ENs are trying to operate within the available resources. We denote the game in MEC node $i$ with $\mathcal{G}_{i}$. We consider a set $\mathcal{A}$ to represent the set of actions (strategies) to be chosen by players, which are given by the VNFs. An action $a_{j} \in \mathcal{A}$ could be a demand for a resource such as CPU, storage, or transmission. The game $\mathcal{G}_{i}$ is given as follows:
\begin{equation}
\label{eq:game}
    \mathcal{G}_{i}=\langle\mathcal{F}_{i},\mathcal{A},(r_j)_{j\in \mathcal{F}_i}\rangle,
\end{equation}
where $r_j$ denotes the payoff function of player $j$. It represents the amount of the allocated resources, given as follows:
\begin{equation}
\label{eq:payoff}
    r_j(\mathcal{A})=\alpha_{i}\times \frac{h(a_j)}{\mathlarger{\sum_{\substack{j'\in\mathcal{F}_i\\j\neq j'}}h(a_{j'})}}-a_j\times\gamma_{(i,j)},
\end{equation}
where $\gamma_{(i,j)}$ is the expected delay of using the resources at the EN $i$ (i.e. $a_j\times\gamma_{(i,j)}$ represents the cost of the service) and $h(\cdot)$ is a mapping that should meet the following conditions:
\begin{equation}
    \begin{cases}
        \mathlarger{\sum_{\substack{v'\in\mathcal{F}_i\\v\neq v'}}}h(a_{j'})>0,\\
        h(0)=0, \text{ and $h(\cdot)$ is a positive function}.
    \end{cases}    
\end{equation}
We define the mean-field term $m_v$, which represents an important parameter in the study of the mean-field game,
\begin{equation}
\label{eq:mf_term}
    m_v=\left(\frac{1}{F}\mathlarger{\sum_{j'\in\mathcal{F}_k}a_{j'}^{\beta}}\right)^{\frac{1}{F}}
\end{equation}
where $\mathcal{F}_k$ is the set of VNFs in the requested SFC. Hence, we can express the payoff function and the mean-field terms differently as follows:
\begin{equation}
\label{eq:r_payoff}
    \begin{cases}
        r_{v}^{*}\left(a_{j},m_{(v,-j)}\right)=\frac{\alpha_i}{v}\left(\frac{a_j}{m_v}\right)^{\beta}-a_j\times\gamma_{(i,j)}\\
        m_{(v,-j)}=\frac{1}{v-1}\mathlarger{\sum}_{\substack{j,j'\in\mathcal{F_i}\\j\neq j'}}a_{j'}^{\beta}
    \end{cases}
\end{equation}
We could then simplify the term of $m_{(v,-j)}$ and rewrite it as:
\begin{equation}
    m_{(v,-j)}=\frac{1}{v-1}\left(m_{v}^{\beta}-\frac{a_{j}^{\beta}}{v}\right)
\end{equation}
where $m_{(v,-j)}$ is the weighted mean when the action of user $j$ is not chosen, and $\alpha_i$ is the amount of resource available of the requested resources. We note here that the payoff depends only on the chosen action and the mean $m_v$. Finally, we rewrite the term of the payoff as:
\begin{equation}
\label{eq:rewri_payoff}
    r_{v}^{*}\left(a_j,m_{(v,-j)}\right)=\frac{a_{j}^{\beta}\times\alpha_i}{m_{(v,-j)}(v-1)+a_{j}^{\beta}}-a_j\times\gamma_{(i,j)}
\end{equation}
In the previous part, we discussed the game within a beforehand known number of VNFs. Nevertheless, the number of VNFs competing over the resources is variable and change over the time depending the number of requests. With increasing number of requests, the number of VNFs to be instantiated increases. Therefore, it is necessary to study this special case of MFG. In this case, we will study the MFG under infinite number of players, which can be performed through applying a slight modification to the payoff function.

Let $\Bar{\mathcal{G}_{i}}=\left\langle\mathcal{A},\Bar{r}_v\right\rangle$ be the game in which the number of players is infinite. A reformulation of the payoff function is required. Thus, we define the new payoff function as follows:
\begin{equation}
\label{eq:refor_payoff}
    \Bar{r}_{v}\left(a_j,m_{(v,-j)}\right)=\begin{cases}
        \frac{a_{j}^{\beta}}{\Bar{m}^\beta}-a_j\times\gamma_{(i,j)}, m>0\\
        0, m=0
    \end{cases}
\end{equation}
The problem of VNF placement and chaining can be defined as the optimal value of the strategy $a_j\in\mathcal{A}$ to be obtained while solving (\ref{eq:refor_payoff}) for infinite number of VNFs and solving (\ref{eq:r_payoff}) for a finite number of VNFs.

\subsection{The VNF scheduling subproblem}

It is shown in \cite{pham2017virtual} that the VNF scheduling problem is NP-complete , thus, cannot be solved in polynomial time, unless $P=NP$. To efficiently solve the problem, we formulate it as a many-to-one matching game where the VNFs and the EN represent players with a list of preferences.

We consider that the access to the EN resource by the VNF starts if and only if the previous VNF for a given SFC is completely finished and forwarded to the following VNF whenever it exists. Let us consider the processing time for each VNF as $p_v$. We also define the starting time of VNF $v$ at EN $i$ as $\tau_{v}^{i}$. The completion time $\xi_{v}^{i}$ is simply the maximum of the sum of the values of $p_v$ and $\tau_{v}^{i}$.

We introduce time related parameters to express the operational assignment between the VNFs and the time slots. Let $x_{(i,v)}^{f,t}$ represents the assignment of time slot $t$ to instance $f$ of VNF $v$ in EN $i$. $\tau_{(i,v)}^{f}$ is the starting time for the forward operation of the instance $f$ of VNF $v$ at the EN $i$, subsequently, $\theta_{(v,t)}^{i}$ is the assignment of time slot $t$ to the forwarding operation.

The objective is to minimize the completion time,

\begin{equation}
    \label{eq:xi}
    \xi_{v}^{i}=\min\left\{\max_{\substack{i\in\mathcal{E}\\v\in\mathcal{F}_i}}\{p_v + \tau_{v}^{i}\}\right\}
\end{equation}
considering the following constraints,
\begin{subequations}
    \begin{equation}
        \label{eq:c1}
        p_v + \tau_{v}^{i} \leq \tau_{v+1}^i
    \end{equation}
    \begin{equation}
        \label{eq:c2}
        \sum_{t=1}^{T} x_{(i,v)}^{f}.x_{(i,v)}^{f,t} \geq x_{(i,v)}^f
    \end{equation}
    \begin{equation}
        \label{eq:c3}
        \tau_{(i,v)}^{f} + \sum_{t=1}^{T} x_{(i,v)}^{f,t}.x_{(i,v)}^{f}\leq \tau_{(v')}^i
    \end{equation}
    \begin{equation}
        \label{eq:c4}
        \tau_{v}^{i} +\sum_{t=1}^{T} x_{(i,v)}^{f}.x_{(i,v)}^{f,t} \leq T_{out}
    \end{equation}
    \begin{equation}
        \label{eq:c5}
        \tau_{(i,v)}^{f} + \sum_{t=1}^{T} x_{(i,v)}^{f,t}.x_{(i,v)}^{f}\leq \tau_{(i,v')}^f
    \end{equation}
    \begin{equation}
        \label{eq:c6}
        \tau_{(i,v)}^{f} + \sum_{t=1}^{T} \theta_{(v,t)}^{i}.y_{z}^{l}\leq \tau_{v+1}^i
    \end{equation}
    \begin{equation}
        \label{eq:c7}
        \sum_{t=1}^{T} x_{(i,v,t)}^{f} + \theta_{(v,t)}^{i} \geq \gamma_{(i,j)}^v
    \end{equation}
\end{subequations}

Constraint in Eq. (\ref{eq:c1}) ensures the order of starting time for the VNF $v$.
Constraint in Eq. (\ref{eq:c2}) ensures the sufficiency of the total allocated time for the VNF $v$.
Constraint in Eq. (\ref{eq:c3}) ensures that the scheduling for the VNF $v$ is not performed earlier than it should be (\ie in cases where VNF $v'$ is waiting to be processed/forwarded).
Constraint in Eq. (\ref{eq:c4}) ensures that the time out of the processing/forwarding of the VNF $v$ is respected.
Constraint in Eq. (\ref{eq:c5}) ensures the completness of the process-forward process (\ie forwarding only the finished VNFs)
Constraint in Eq. (\ref{eq:c6}) similar to Eq. (\ref{eq:c5}) ensures that a VNF will not be processed until the total forward of the previous VNF.
Constraint in Eq. (\ref{eq:c7}) ensures that the requirement of each VNF are met.

%% file: proposed_solution.tex
\section{Theoretical Game Approach solution}
In this section, we will present our approach for MEC-enabled SFC resource allocation. The VNFs placement and chaining is provided through the resolution of the MFG. Then, based on the placement and chaining provided by the previous step, we provide a matching-based solution to schedule the VNFs' execution on different ENs.

\subsection{The VNFs Placement and Chaining}

\subsubsection{Mean-Field Game based resource provisioning}

\begin{algorithm}[!b]
\SetAlgoLined
\label{algo:imla}
    \KwIn{$\lambda$, $\mathcal{A}$, $\beta$, $\rho_{j,k}$}
    
    \nonl \textbf{\textit{Learning stage}}\;
    \For{$t=0$, $t<iterations$}{
        Choose random action $a_j \in \mathcal{A}$\;
        Compute $BRC(a_j)$ from Eq. (\ref{eq:BRC})\;
        Update the mean term in Eq. (\ref{eq:mf_term})
    }
    
    \nonl \textbf{\textit{Placement stage}}\;
    Choose VNFs with optimal actions according to Eq. (\ref{eq:argmax})\;
    Instantiate the VNFs containers\;
    
    \nonl\textbf{\textit{Routing stage}}\;
    Get other nodes VNFs placement\;
    Select next VNFs for the requested SFC\;
    Solve the problem of Eq. (\ref{eq:xi}) to forward processed packets\;
 \caption{IMLA: Ishikawa-Mann Learning Algorithm.}
\end{algorithm}
The MFG-based solution relay on the IMLA algorithm (Algorithm \ref{algo:imla}). IMLA is performed on three stages; the learning stage, the placement stage and the routing stage.
At the learning stage, we train our model to find the optimal strategy for each VNF. At each iteration, the controller chooses a random action, then compute the best response correspondence solution from Eq. (\ref{eq:BRC}). Then it stores the obtained values, and updates the mean-field term according to Eq. (\ref{eq:mf_term}). At the end of this stage, we obtain the optimal strategy that makes the system converges to a Nash equilibrium. To place the VNFs accordingly, we choose the requested ones having the optimal actions, then instantiate the adequate VNFs' containers. Finally, regarding the routing stage, the SDN controller gets the placement of the VNFs on other nodes. Therefore, each VNF forwards the processed packets to the next VNF on the SFC.
\subsubsection{Equilibrium analysis}
Let us first introduce the best response correspondence problem (BRC) for each player (VNF in our case) when we have a fixed number of players. BRC is presented as follows; given the set of all possible strategies of all players but $j$, $a_{-j}=\{a_1,\ldots,a_{j-1},a_{j+1},\ldots\}$, find the maximum of their payoff functions, i.e. find $a_j$ as:

\begin{equation}
    \label{eq:argmax}
    a_j\in\argmax_{a_{j'}}\left\{r_j(a_{j'}, a_{-j})\right\}
\end{equation}
where $a_j$ in this problem represents the best response of player $j$ when another player asks for the same type of resources. BRC have a solution if we obtain a negative second-order derivative of the payoff $r_{v}^{*} (a_j,m_{(v,-j)})$; $\partial_{a_j}^{2} r_{v}^{*} (a_j,m_{(v,-j)})<0$ (see appendix \ref{app:bcp-deriv} for the proof), therefore, a Nash equilibrium exists and is given by:
\begin{equation}
    \sqrt{
        a_{j}^{\beta-1}\left(\frac{\beta\rho_i}{v\gamma_{(i,j)}}\frac{1}{v}
        \mathlarger{\sum_{\substack{j,j'\in\mathcal{F}_i\\j\neq j'}}}
        a_{j'}^{\beta}\right)
    }-\frac{a_{j}^{\beta}}{v}-\frac{1}{v}\sum_{\substack{j,j'\in\mathcal{F}_i\\j\neq j'}}a_{j'}^{\beta}=0
\end{equation}
We notice that the actions space and the payoff function are identical and invariant respectively, which means that this game is symmetric\cite{hanif2015mean}. Therefore, NE is given as follows:
\begin{equation}
    \label{eq:sym_game}
    \sqrt{
        a_{j}^{\beta-1}\left(\frac{\beta\rho_i}{v\gamma_{(i,j)}}\frac{v-1}{v}a_{j'}^{\beta}\right)
    }-\frac{a_{j}^{\beta}}{v}-\frac{v-1}{v}a_{j'}^{\beta}=0
\end{equation}
By developing the term of Eq. (\ref{eq:sym_game}), we obtain the NE value for the BRC:
\begin{equation}
    a_{j}^{*}=\frac{\beta\rho_i\left(v-1\right)}{v^2\gamma_{(i,j)}}
\end{equation}
In addition, whenever a equilibrium exists, all players have symmetric strategies, and, the NE payoff is given as,
\begin{equation}
    r_{v}^{*}(a_{j}^{*})=a_j\gamma_{(i,j)}\left[\frac{a_{j}^{\beta}+M}{\beta M}-1\right]
\end{equation}
with $M$ is given as,
\begin{equation}
    M=\frac{1}{v}\mathlarger{\sum_{\substack{j'\in\mathcal{F}_i\\j\neq j'}}a_{j'}^{\beta}}
\end{equation}
For values of $\beta\in[0,1]$, we have $r_{v}^{*}\geq0$, otherwise, if $\beta>1$, then the strategies of the players using the resources depends on others that does not use the resources and the number of VNFs executing tasks is satisfying:
\begin{equation}
    \frac{\beta}{\beta-1}\geq v
\end{equation}
Under these conditions, the solution of BRC is equal to $0$. The resources are not wasted (\ie efficient usage) when the delay is equal to,
\begin{equation}
    \gamma_{(i,j)}=\beta\times\frac{v-1}{v}
\end{equation}
For $\gamma_{(i,j)}<\beta$, the requested amount of resources and the available resources are equal at the equilibrium, hence, the efficient ratio is written as
\begin{equation}
    \nu=v\times\frac{a_{NE}^{*}}{\rho_i}
\end{equation}
The equilibrium increases with $\beta$ and the amount of requested resources and decreases with the delay $\gamma_{(i,j)}$.

\subsubsection{Reinforcement Learning based solution}
In this paper we propose an iterative learning algorithm that converges to a NE. This algorithm is executed by the SDN controller to compute the adequate placement and chaining of the VNFs in the SFCs. The benefit of using an iterative algorithm is the low computational complexity required for reaching the equilibrium. We use the following formulation based on Ishikawa-Mann iteration \cite{ishikawa1974fixed},
\begin{equation}
    a_{(t+1)}=\lambda BRC(a_t) + a_{t}(1-\lambda)
\end{equation}
where $\lambda$ is the learning rate, and BRC is the solution function to the problem in Eq. (\ref{eq:argmax}), given as follows,
\begin{equation}
    \label{eq:BRC}
    BRC(a_j)=\max\left\{v\sqrt{\frac{\beta\alpha_i}{v\gamma_i}\left(m_{}-\frac{a_{}}{v}\right)}-\left(m_{}-\frac{a_{}}{v}\right), 0\right\}
\end{equation}
\\
Eq. (\ref{eq:BRC}) is obtained through solving the equation $BRC(a_{j}^{*})=a_{j}^{*}$.

Regarding the infinit regime (\ie cases with huge number of VNFs competing over the EN resources), we stick with the same solution approach based on the BRC, with a small change regarding the mean-field term. In the case of huge number of users, the mean-field term is also updated accordingly with the evolution of the BRC stages as follows,
\begin{equation}
    a_{\left(i,t+1\right)}\in\argmax_{{a'}_{j}}\left\{\Bar{r}(a_{j}^{'}, m_{t})\right\}
\end{equation}
and the main objective is to maximize the payoff at the next stage, given as,
\begin{equation}
    m_{t+1}=\lambda.BRC(m_t) + m_t(1-\lambda)
\end{equation}

The complexity of the proposed Ishikawa-Mann algorithm (Algorithm \ref{algo:imla}) depends on the complexity of the BRC function. However, considering that the values of the learning rate are smaller and fixed, the BRC converges to a fixed point $\iota$ in near-proximity of the equilibrium value $a_{j}^{*}$, which makes the convergence time $\log(\frac{1}{\iota})$ \cite{hanif2015mean}. Therefore, we conclude that proposed approach converges in a logarithmic time $O(\log(v))$.

\subsection{The VNF scheduling subproblem}
\begin{figure}[!b]
    \centering
	\includegraphics[width=.8\linewidth]{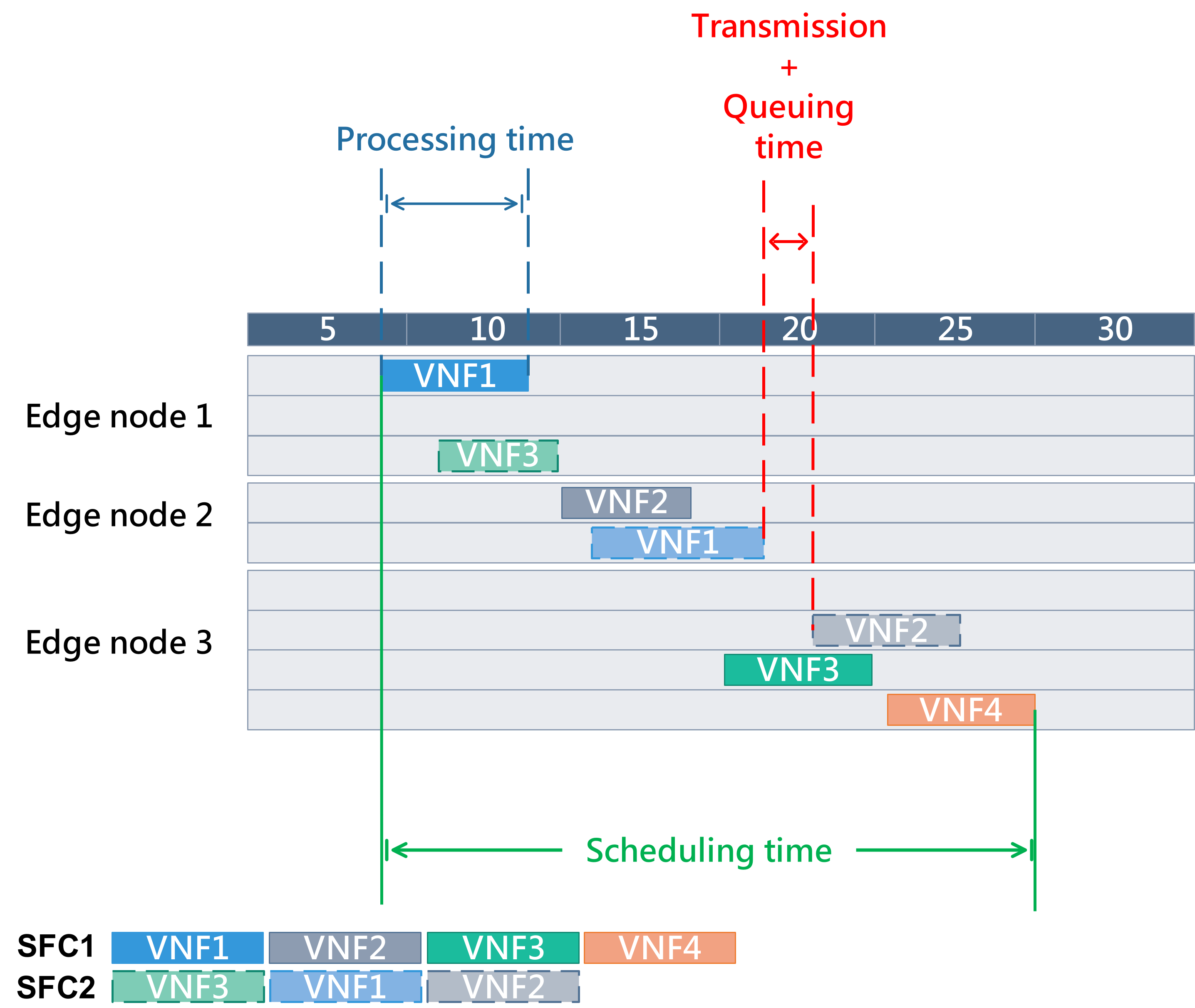}
    
    \caption{\label{fig:empl-sched}Scheduling time example for the SFC presented in Fig. \ref{fig:exmpl-sfc}.}
\end{figure}
TO illustrate the subproblem, Fig. \ref{fig:empl-sched} provides an example of SFC scheduling based on the example of Fig. \ref{fig:exmpl-sfc}. The scheduling time is essentially the sum of the processing time for each VNF in the SFCs, the transmission and the queuing delay. By processing we mean the processing of the request by a given VNF, which could include the computational and the storage. The transmission time is considered to take part of the queuing part.

\subsubsection{Many-To-One Matching Game}
Considering the constraints on resource provisioning and the execution order of VNFs from the previous section, the SDN controller, as a supervising entity of the network, needs to find the adequate schedule to execute the VNFs of the SFCs on the different nodes of the system. The VNFs-resources assignment at each moment of the network operations is considered as the output of a many-to-one matching game. The players are the VNFs showing their interest to be matched to the resources of an EN and we define the requirements for a matching game, namely, the stability, blocking pairs, and preferences lists of each set of players as follows. For simplicity, the VNFs-resources assignment will be denoted as VFNs-ENs assignment.

\SetKwInOut{Initialization}{Initialization}
\begin{algorithm}[!t]
\SetAlgoLined
\label{algo:emsda}
    \KwIn{$\succ_{f_{(i,j)}}$, $\mathrm{Fp}$, $\succ_{i}$, $F_k$, $E$, $q_{i}^{min}$, $q_{i}^{max}$}
    \KwOut{$\mu(f_{(i,j)})$}
    
    \Initialization {$q_{i}^{0,min}=q_{i}^{min}$, $q_{i}^{0,max}=q_{i}^{max}$, $\forall i \in \mathcal{E}$}
    
    \While{$\mathrm{Fp}\neq\emptyset$}{
        $\mathrm{Q}^\theta=\mathlarger{\sum}_{i\in\mathcal{E}}q_{i}^{\theta,min}$\;
        $R^{\theta}=\{f_v, \dots\}$ \text{ with } $\sum_{i\in \mathcal{E}}\alpha_i(t) \leq Q$\;
        $UM=R^{\theta-1} \backslash R^{\theta}$\;
        \eIf {$UM\neq\emptyset$}{
            Apply classic DAA on unmatched VNFs in $UM$ with $q_{i}^{\theta,max}$\;
        }{
            Apply classic DAA on unmatched VNFs in $UM$ with $q_{i}^{\theta,min}$\;
        }
        \If{(\ref{eq:c3})-(\ref{eq:c5}) \text{are met}}{
            Add matched VNFs to $\mu$\;
            Remove matched VNFs from $\mathrm{Fp}$\;
            Update the quotas according to Eqs. (\ref{eq:update_q_min}) and (\ref{eq:update_q_max})\;
        }
    }
    \Return $\mu$ 
    \caption{eMSDA: enhanced Multi-Staged Deferred Acceptance Algorithm.}
\end{algorithm}

The preference is defined as binary relationship defined between the elements of the players sets. The relationship is  complete, reflexive, and transitive. Let us denote $\succ_i$ the preferences list of the ENs and $\succ_{f_{(i,j)}}$ the preference list of the VNFs. $e_i \succ_{f_{(i,j)}} e_{i'}$ means that the VNF $f_{(i,j)}$ prefers the EN $i$ over the EN $e_{i'}$. Using the same notation, we define the preferences of ENs over VNFs by $\succ_i$.
\begin{definition}[\textbf{Many-to-One matching game}]
A many-to-one matching game is a two-sided assignment problem between two disjoint sets of players where the players on the first set express their interest to be matched to a player of the other set based on a preference.
Let $\mu$ be a many-to-one mapping from the set of VNFs to the set of ENs. $\mu:\mathcal{F}_i\mapsto\mathcal{E}$ satisfying the following conditions:
\begin{itemize}
    \item $\forall f_{(i,j)} \in \mathcal{F}_i$ we have $\mu(f_{(i,j)}) \in \mathcal{E}$
    \item $\forall e_i$ we have $\mu(e_i) \in \mathcal{F}_i$
    \item $\mu(f_{(i,j)})=e_i$ exits, only if $e_i\in\mathcal{F}_i$
\end{itemize}
We consider that the VNFs-EN assignment takes into account a minimum and maximum number of VNFs to be matched to an EN to operate. These quantities are called quotas (minimal and maximal) and ensure a good resource consumption and fairness. Let $q_{i}^{max}$ and $q_{i}^{min}$ be the maximum and minimum quota, respectively, of the EN $i$. $\mu$ is feasible if and only if $q_{i}^{min} \leq q_i \leq q_{i}^{max} $ with, $q_i=|\mu(f_{(i,j)})|$ and $|\mu(f_{(i,j)})|=\{0,1\}$. A stable matching is defined as follows:
\end{definition}
\begin{definition}[\textbf{Stable matching}]
    A matching $\mu$ is said to be stable if there is no intention from any pair $(f_{(i,j)}, e_i)$ to deviate from $\mu$. In other words there is no blocking pair is defined as follows:
\end{definition}

\begin{definition}[\textbf{Blocking pair}]
    $(f_{(i,j)}, e_i)$ is said to be a blocking pair if it satisfies the following conditions:
    \begin{itemize}
        \item $e_i\succ_{f_(i,j)}e_{i^{'}}$ , for $e_{i^{'}}\in\mu(f_{(i,j)})$
        \item $f_{(i,j)}\succ_i f'_{(i,j)}$
    \end{itemize}
    Under the previous two definitions, $\mu$ is stable and guarantee stability and fairness.
\end{definition}

\textit{a - The VNFs preference List}\\
As for the preferences of the VNFs, they always prefer an EN that will process/forward them in the shortest delay, hence, with available resources (based on the results of the VNFs placement and chaining) at the current time slot. The preference list of VNFs is given as follows,
\begin{equation}
    \tau_{f_{(i,j)}}(a_j)=t
\end{equation}

\textit{b - The EN Preference List}\\
The edge nodes, however, are more interested in executing/forwarding VFNs with smaller processing requirements.
\begin{equation}
    \tau_{a_{i}}(f_{(i,j)})=\inf_{j\in\mathcal{F}_i}\left[\rho_{j}\right]
\end{equation}

Considering the quota constraints, the classic DAA \cite{gale2013college} will not satisfy the feasibility of the matching (see appendix \ref{app:min-quota}). We consider a staged deferred acceptance algorithm that support the constraints of minimum and maximum quota. Our algorithm is called enhanced multi-stage deferred acceptance algorithm (eMSDA) and detailed in Algorithm \ref{algo:emsda}. In addition, Nash equilibrium exists whenever a stable matching exists \cite{han2012game}.

\subsubsection{Enhanced Multistage Deferred Acceptance Algorithm}
eMSDA is developed in Algorithm \ref{algo:emsda} inspired from the work in \cite{fragiadakis2016strategyproof}, which is based on a DAA algorithm performed in several stages. eMSDA works under the assumption that all the ENs share the same preference list ranking all the VNFs following their requirements in terms of resources. In addition, since the network is controlled by the SDN controller, this ensures that that the process of making the preferences list easy. We denote by $\mathrm{Fp}$ the list of preferences, and $\succ_{\mathrm{Fp}}$ the shared preference relationship. Initially, we temporarily reserve a subgroup of VNFs and perform the DAA on the remaining subgroup. The assignments at a given stage $\theta$ for a subset of EN and VNFs are considered final, then accordingly, we reduce the minimum and maximum quotas as, 
\begin{subequations}
    \begin{equation}
    \label{eq:update_q_min}
        q_{i}^{\theta,min}=max\left\{q_{i}^{\theta-1,min}\mathlarger{\sum_{i\in\mathcal{E}}}\alpha_{i}(t), 0\right\}
    \end{equation}
    \begin{equation}
    \label{eq:update_q_max}
        q_{i}^{\theta,max}=q_{i}^{\theta,max}-\mathlarger{\sum_{i\in\mathcal{E}}}\alpha_{i}(t)
    \end{equation}
\end{subequations}
At every stage we rank the VNFs into the most preferred and the least preferred. In fact, the number of the least preferred VNFs is the sum of the minimum quota of all the ENs. Depending on the number of preferred and non-preferred, we run the DAA with maximum quota. Moreover, VNFs cannot be matched unless the previous VNF is finished.

%% file: simulations.tex
\section{Simulation Results}
\begin{figure}[!t]
    \centering
    \includegraphics[width=.8\linewidth]{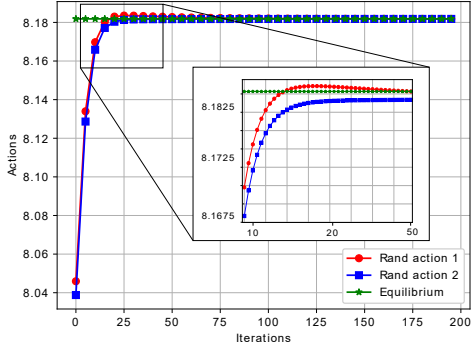}
    \caption{MFG Stability for finite number of VNFs. $\lambda=0.05$, $v=10$.}
    \label{fig:nash_rand_action_finit}
\end{figure}

\begin{figure}[!t]
    \centering
    \includegraphics[width=.8\linewidth]{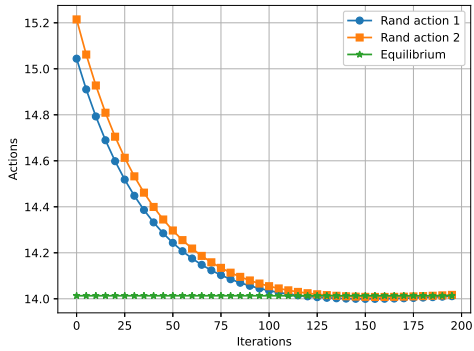}
    \caption{MFG Stability for infinite number of VNFs. $\lambda=0.05$, $v=1000$.}
    \label{fig:nash_rand_action_infinit}
\end{figure}

In this section, we present the simulation results to illustrate the performance of the proposed IMLA for the VNF placement and chaining subproblem, and the proposed eMSDA for the VNFs scheduling subproblem.
We compare the performance of the proposed algorithms in terms of execution delay and resource consumption to the performance of a genetic algorithm based solution and an integer linear programming algorithm.
\subsection{Games Stability and Convergence}
To prove the existence of the equilibrium for the VNF placement and chaining problem, we simulated the BRC solution and compared it to the NE obtained mathematically. We performed exhaustive simulations on both finite and infinite regimes. Results are illustrated in Fig. \ref{fig:nash_rand_action_finit} and Fig. \ref{fig:nash_rand_action_infinit}. For both scenarios, the actions sets were generated randomly as, $a_j\in\mathcal{A} \text{ and } \mathcal{A}=\mathbb{R}_+$.
Fig. \ref{fig:nash_rand_action_finit} illustrates the evolution of the chosen action when applying the IMLA algorithm. It is shown that both generated actions sets converges to a NE, and reached it approximately after 35 iterations of learning time.

IMLA algorithm converges to a NE whenever the learning rate is sufficiently small and constant. In fact, the function in Eq. (\ref{eq:BRC}) converges to a fixed point (i.e. equilibrium) with small values of $\lambda$, precisely when $\lambda\in(0,1)$ \cite{hanif2015mean}. In our case, we set $\lambda$ equals to $0.05$.

\subsection{System Evaluation}
In this section, we present the performance of the proposed approaches (IMLA) and (eMSDA) in terms of operations delay, execution delay, and resource consumption. In order to evaluate the performance of the algorithms, we compared them to benchmarks of existing approaches from the literature. In particular, we compared the approach to an meta-heuristic-based approach, and precisely the genetic algorithm (GA) \cite{8892907}, and integer linear-programming (ILP) \cite{9060929}.
The simulation setup consists in variable set of EN generated with values between $3$ and $15$, varying VNFs between $1$ and $1000$ per EN. The requests are generated randomly, so are the requirement of each EN. We fixed the value of $\beta=1$, $\lambda=0.05$, the learning iterations in $200$, and the packet's size between $100$ and $10000$ Kb.

%
%
\begin{figure}[!t]
    \centering
    \includegraphics[width=.8\linewidth]{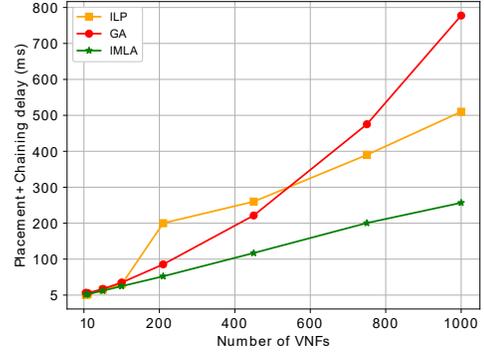}
    \caption{Placement and chaining delay in function of the packet size.}
    \label{fig:vnf_delay_pc}
\end{figure}

\begin{figure}[!t]
    \centering
    \includegraphics[width=.8\linewidth]{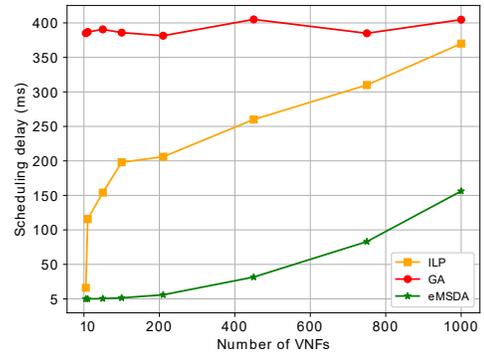}
     \caption{Scheduling delay in function of the number of VNFs.}
    \label{fig:vnf_delay_schd}
\end{figure}

Fig. \ref{fig:vnf_delay_pc} and Fig. \ref{fig:vnf_delay_schd} illustrate the performance of the proposed approach in terms of execution delay as a function of the number of the VNFs. It is shown that the proposed algorithms outperforms both of other approaches. In fact, both of the proposed algorithms (\ie IMLA and eMSDA) have, first, a lower complexity compared to other algorithms in terms of their implementations. Second,  for instance the IMLA execution time depends on the convergence of the BRC solution, which is very low compared to the benchmark algorithms (\ie $O(\log(v))$ vs $O(Mv^2)$, with $M$ is the number of generations after which the GA converges to a solution). As for the scheduling, the GA approach show a constant behavior compared to the proposed approach and to the ILP-based approach. Although a stable behavior when it comes to a huge number of VNFs, it is not beneficial in case the number of VNFs to schedule is small. Fig. \ref{fig:vnf_delay_pc} and Fig. \ref{fig:vnf_delay_schd} show that the proposed approach based on the game theory model consumes less time for the operational delay, nearly $40\%$ less than the GA and $25\%$ compared to ILP in the case where the number of VNFs to manage is sufficiently big (\ie ~1000 in this case).
%
%

\begin{figure}[!t]
    \centering
    \includegraphics[width=.8\linewidth]{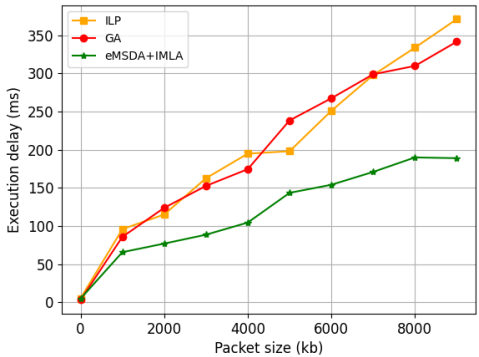}
    \caption{Processing time in function of the packet size.}
    \label{fig:proc_time}
\end{figure}

Fig. \ref{fig:proc_time} shows the evolution of the delay as a function of the packet's size for both the eMSDA and the IMLA. In this scenario, we considered the average delay obtained from the different setups of the random scenarios through varying the number of ENs, SFCs, and the number of VNFs per SFC and also varying the size of the packets processed by the SFCs. In most of the cases, the proposed approach shows an enhacement in terms of execution delay of the SFCs. The proposed approach offers a reduced delay (around $45\%$) compared to other approaches when it comes to processing different configurations of packets in terms of packet's size. Such result is justified by, first, the reduced time that both proposed algorithms are showing in terms of placing and chaining, and second, by the enhanced scheduling provided by the eMSDA, that not only provides a stable matching for the VNFs and ENs resources but also provides a low delay in terms of execution time.
%
%

\begin{figure}[!t]
    \centering
    \includegraphics[width=.8\linewidth]{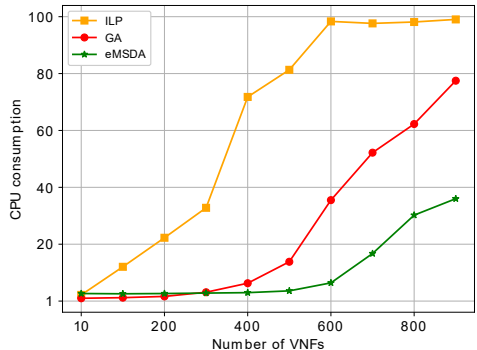}
     \caption{CPU consumption evolution in function of the number of VNFs.}
    \label{fig:cpu_consump}
\end{figure}

\begin{figure}[!t]
    \centering
    \includegraphics[width=.8\linewidth]{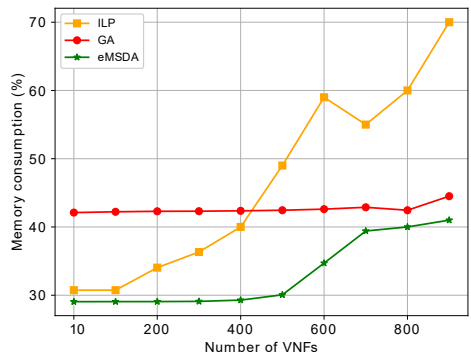}
     \caption{Memory consumption evolution in function of the number of VNFs.}
    \label{fig:memory_consump}
\end{figure}

Fig. \ref{fig:cpu_consump} and Fig. \ref{fig:memory_consump} illustrate the resource consumption of the proposed algorithm compared to the GA and ILP approaches. It is shown that the proposed approach gives better results in terms of resource consumption (\ie CPU and memory). The GA approach shows a constant behavior also in terms of memory consumption, which can explain more the results obtained in Fig. \ref{fig:vnf_delay_schd}. As for ILP, the consumption of the CPU and the memory is still higher than the proposed approach and the GA. We can conclude from Fig. \ref{fig:cpu_consump} and Fig. \ref{fig:memory_consump} that the proposed approach offers enhanced results compared the benchmarked approaches. The theoretical game based approach shows around $40\%$ lower than the ILP approach and $25\%$ better than GA  especially when the number of VNFs is large.
%
%
%
%

%% file: related_works.tex
\section{Related Works}
The VNF/SFC resource allocation approaches can be grouped into three main categories, theoretical games approaches \cite{pham2017virtual,abouaomar2018users}, linear programming variant approaches \cite{9060929,9181472,9013429}, the machine learning approaches \cite{9075271,9187253} and the heuristic approaches \cite{8647858,9062531}.

Work in \cite{pham2017virtual} highlighted the importance of the VNF scheduling, and proposed a matching-based algorithm, namely, the one-to-one matching to cope with the scheduling problem which is NP-hard. The proposed approach guarantees the stability of the scheduling process. However, the authors in this work consider only the scheduling of one VNF at a time, and even in the transmission phase, the whole channel is allocated to one and only one VNF. Compared to our work, we considered a many-to-one matching game to schedule multiple VNFs at a time.
Work in \cite{9060929} proposed an online provisioning for NFV considering both unicast and multicast services with two types of VNFs, mandatory and best-effort instances. In addition, the work considered only the computational and transmission resources. Yet, it did not make any assumption about the resource state at the edge nodes and consider only two types of resources. The same work also supposed that the requests arrives in batches having the same requirements of resources which may not cover all the use cases of SFC resource provisioning while placing and chaining the VNFs. The authors also did not consider the scheduling problem and only investigated the routing and placement of the VNFs. 
Work in \cite{9181472} addressed the problem of reliability, delay, and resource allocation for SFC in softwarized 5G networks. The work suggested an SFC sub-chaining method to cope with the reliability and an ILP based solution is proposed for the problem of SFC placement through a matching algorithm that provides a new-optimal polynomial solution. However, in this work, the authors did not provide much detail about the format of the request, types of resources, nor the dynamic of request arrival. In addition, the proposed matching algorithm might not be efficient when the maximum capacities of the nodes are considered.
For the machine learning approaches, the authors in \cite{9075271} investigated the problem of VNFs scheduling with the aim to minimize the overall completion time of the services under delay constraints. Authors proposed a RL based solution to learn the best scheduling scheme. In fact, the problem of scheduling is formulated as a mixed integer linear program and is relaxed as an Markov decision process and solved using RL. However, the request arrival rate in this work does not considered stochastic and the RL is based on Q-Learning algorithm, which acts only on a set of specific fixed instance of the problem. Work in \cite{9187253} models the SFC scheduling problem as a flexible job-shop scheduling problem with the objective to minimize the scheduling latency. Authors proposed a deep RL based on Q-Learning that gives the environment the advantage of performing adaptive scheduling. 

Within a heuristic context, work in \cite{8892907} investigated the problem of resource scheduling with the aim to enhance the usage of network resources as well as the minimization of the experienced end-to-end delay for the network services. Authors proposed an approach based on the GA through improving the crossover and mutation operations. Although GAs are good in supporting multi-objective optimization, the cost of computation in terms of time is very high. Within the same context, authors in \cite{9062531} proposed a configurable service allocation scheme for VNF embedding and routing. Due the NP-hardness of the investigated problem, the work is considered as an integer non-linear programming model and solved it through heuristic methods, specifically a greedy algorithm. The proposed solution leveraged the properties of the different entities of the system and balanced the resources consumption. Work in \cite{8647858} proposed a VNF placement scheme for MEC environment and formulated of VNF placement problem under minimizing access latency and maximizing service availability constraints. To reduce the problem complexity, the authors proposed a GA and compared the obtained results with a CPLEX implementation.

In our work, we first, consider four types of resources, namely, computation, transmission, and memory. Secondly, we consider a stochastic request arrival model which mimic as real as possible a real SFC deployment. Finally, we consider that the VNFs might have different resource requirement and we tackled the problem of resource allocation considering the VNFs placement, chaining, and scheduling.

%% file: bare_jrnl.bbl
\begin{thebibliography}{10}
\providecommand{\url}[1]{#1}
\csname url@samestyle\endcsname
\providecommand{\newblock}{\relax}
\providecommand{\bibinfo}[2]{#2}
\providecommand{\BIBentrySTDinterwordspacing}{\spaceskip=0pt\relax}
\providecommand{\BIBentryALTinterwordstretchfactor}{4}
\providecommand{\BIBentryALTinterwordspacing}{\spaceskip=\fontdimen2\font plus
\BIBentryALTinterwordstretchfactor\fontdimen3\font minus
  \fontdimen4\font\relax}
\providecommand{\BIBforeignlanguage}[2]{{%
\expandafter\ifx\csname l@#1\endcsname\relax
\typeout{** WARNING: IEEEtran.bst: No hyphenation pattern has been}%
\typeout{** loaded for the language `#1'. Using the pattern for}%
\typeout{** the default language instead.}%
\else
\language=\csname l@#1\endcsname
\fi
#2}}
\providecommand{\BIBdecl}{\relax}
\BIBdecl

\bibitem{8985528}
J.~{Navarro-Ortiz} \emph{et~al.}, ``A survey on 5g usage scenarios and traffic
  models,'' \emph{IEEE Communications Surveys Tutorials}, vol.~22, no.~2, pp.
  905--929, 2020.

\bibitem{9055745}
A.~{Ksentini} \emph{et~al.}, ``Toward slicing-enabled multi-access edge
  computing in 5g,'' \emph{IEEE Network}, vol.~34, no.~2, pp. 99--105, 2020.

\bibitem{filali2020multi}
A.~Filali \emph{et~al.}, ``Multi-access edge computing: A survey,'' \emph{IEEE
  Access}, 2020.

\bibitem{nour2020computeless}
B.~Nour \emph{et~al.}, ``{Compute-Less Networking: Perspectives, Challenges,
  and Opportunities},'' \emph{IEEE Network}, vol.~34, no.~6, pp. 259--265,
  2020.

\bibitem{9263348}
A.~Filali \emph{et~al.}, ``Preemptive sdn load balancing with machine learning
  for delay sensitive applications,'' \emph{IEEE Transactions on Vehicular
  Technology}, vol.~69, no.~12, pp. 15\,947--15\,963, 2020.

\bibitem{7534741}
I.~Alam \emph{et~al.}, ``A survey of network virtualization techniques for
  internet of things using sdn and nfv,'' \emph{ACM Computing Surveys (CSUR)},
  vol.~53, no.~2, pp. 1--40, 2020.

\bibitem{abouaomar2017caching}
A.~Abouaomar \emph{et~al.}, ``Caching, device-to-device and fog computing in 5
  th cellular networks generation: Survey,'' in \emph{International Conference
  on Wireless Networks and Mobile Communications (WINCOM)}.\hskip 1em plus
  0.5em minus 0.4em\relax IEEE, 2017, pp. 1--6.

\bibitem{tak2020federated}
A.~Tak \emph{et~al.}, ``Federated edge learning: Design issues and
  challenges,'' \emph{IEEE Network}, 2020.

\bibitem{pham2017virtual}
C.~Pham \emph{et~al.}, ``Virtual network function scheduling: A matching game
  approach,'' \emph{IEEE Communications Letters}, vol.~22, no.~1, pp. 69--72,
  2017.

\bibitem{9060929}
O.~{Alhussein} \emph{et~al.}, ``Robust online composition, routing and nf
  placement for nfv-enabled services,'' \emph{IEEE Journal on Selected Areas in
  Communications}, vol.~38, no.~6, pp. 1089--1101, 2020.

\bibitem{9181472}
P.~{KaliyammalThiruvasagam} \emph{et~al.}, ``A reliability-aware, delay
  guaranteed, and resource efficient placement of service function chains in
  softwarized 5g networks,'' \emph{IEEE Transactions on Cloud Computing}, pp.
  1--1, 2020.

\bibitem{9013429}
J.~{Li} \emph{et~al.}, ``On dynamic mapping and scheduling of service function
  chains in sdn/nfv-enabled networks,'' in \emph{IEEE Global Communications
  Conference (GLOBECOM)}, 2019, pp. 1--6.

\bibitem{9075271}
J.~{Li} \emph{et~al.}, ``Delay-aware vnf scheduling: A reinforcement learning
  approach with variable action set,'' \emph{IEEE Transactions on Cognitive
  Communications and Networking}, pp. 1--1, 2020.

\bibitem{9187253}
T.~{Wang} \emph{et~al.}, ``Adaptive service function chain scheduling in mobile
  edge computing via deep reinforcement learning,'' \emph{IEEE Access}, vol.~8,
  pp. 164\,922--164\,935, 2020.

\bibitem{8647858}
L.~{Yala} \emph{et~al.}, ``Latency and availability driven vnf placement in a
  mec-nfv environment,'' in \emph{IEEE Global Communications Conference
  (GLOBECOM)}, 2018, pp. 1--7.

\bibitem{9062531}
G.~{Wang} \emph{et~al.}, ``Sfc-based service provisioning for reconfigurable
  space-air-ground integrated networks,'' \emph{IEEE Journal on Selected Areas
  in Communications}, vol.~38, no.~7, pp. 1478--1489, 2020.

\bibitem{8892907}
Q.~{Li} \emph{et~al.}, ``An improved genetic algorithm for the scheduling of
  virtual network functions,'' in \emph{20th Asia-Pacific Network Operations
  and Management Symposium (APNOMS)}, 2019, pp. 1--4.

\bibitem{lhazmir2019performance}
S.~Lhazmir \emph{et~al.}, ``Performance analysis of uav-assisted ferrying for
  the internet of things,'' in \emph{IEEE Symposium on Computers and
  Communications (ISCC)}.\hskip 1em plus 0.5em minus 0.4em\relax IEEE, 2019,
  pp. 1--6.

\bibitem{azizian2017vehicle}
M.~Azizian \emph{et~al.}, ``Vehicle software updates distribution with sdn and
  cloud computing,'' \emph{IEEE Communications Magazine}, vol.~55, no.~8, pp.
  74--79, 2017.

\bibitem{abouaomar2019resources}
A.~Abouaomar \emph{et~al.}, ``A resources representation for resource
  allocation in fog computing networks,'' in \emph{IEEE Global Communications
  Conference (GLOBECOM)}.\hskip 1em plus 0.5em minus 0.4em\relax IEEE, 2019,
  pp. 1--6.

\bibitem{9326402}
A.~{Abouaomar} \emph{et~al.}, ``Resource provisioning in edge computing for
  latency sensitive applications,'' \emph{IEEE Internet of Things Journal}, pp.
  1--1, 2021.

\bibitem{abouaomar2018matching}
A.~Abouaomar \emph{et~al.}, ``Matching-game for user-fog assignment,'' in
  \emph{IEEE Global Communications Conference (GLOBECOM)}.\hskip 1em plus 0.5em
  minus 0.4em\relax IEEE, 2018, pp. 1--6.

\bibitem{wolff1982poisson}
R.~W. Wolff, ``Poisson arrivals see time averages,'' \emph{Operations
  Research}, vol.~30, no.~2, pp. 223--231, 1982.

\bibitem{gopalan2006statistical}
K.~Gopalan \emph{et~al.}, ``Statistical admission control using delay
  distribution measurements,'' \emph{ACM Transactions on Multimedia Computing,
  Communications, and Applications (TOMM)}, vol.~2, no.~4, pp. 258--281, 2006.

\bibitem{marin2001statistical}
G.~A. Marin \emph{et~al.}, ``Statistical call admission control,'' Apr.~24
  2001, uS Patent 6,222,824.

\bibitem{gupta2019stochastic}
A.~Gupta \emph{et~al.}, ``Stochastic online metric matching,'' \emph{arXiv
  preprint arXiv:1904.09284}, 2019.

\bibitem{dehghani2017stochastic}
S.~Dehghani \emph{et~al.}, ``Stochastic k-server: How should uber work?''
  \emph{arXiv preprint arXiv:1705.05755}, 2017.

\bibitem{thakral2019matching}
N.~Thakral, ``Matching with stochastic arrival,'' in \emph{AEA Papers and
  Proceedings}, vol. 109, 2019, pp. 209--12.

\bibitem{caines2015mean}
P.~E. Caines \emph{et~al.}, ``Mean field games.'' 2015.

\bibitem{hanif2015mean}
A.~F. Hanif \emph{et~al.}, ``Mean-field games for resource sharing in
  cloud-based networks,'' \emph{IEEE/ACM Transactions on Networking}, vol.~24,
  no.~1, pp. 624--637, 2015.

\bibitem{ishikawa1974fixed}
S.~Ishikawa, ``Fixed points by a new iteration method,'' \emph{Proceedings of
  the American Mathematical Society}, vol.~44, no.~1, pp. 147--150, 1974.

\bibitem{gale2013college}
D.~Gale \emph{et~al.}, ``College admissions and the stability of marriage,''
  \emph{The American Mathematical Monthly}, vol. 120, no.~5, pp. 386--391,
  2013.

\bibitem{han2012game}
Z.~Han \emph{et~al.}, \emph{Game theory in wireless and communication networks:
  theory, models, and applications}.\hskip 1em plus 0.5em minus 0.4em\relax
  Cambridge university press, 2012.

\bibitem{fragiadakis2016strategyproof}
D.~Fragiadakis \emph{et~al.}, ``Strategyproof matching with minimum quotas,''
  \emph{ACM Transactions on Economics and Computation (TEAC)}, vol.~4, no.~1,
  pp. 1--40, 2016.

\bibitem{abouaomar2018users}
A.~Abouaomar \emph{et~al.}, ``Users-fogs association within a cache context in
  5g networks: Coalition game model,'' in \emph{2018 IEEE Symposium on
  Computers and Communications (ISCC)}.\hskip 1em plus 0.5em minus 0.4em\relax
  IEEE, 2018, pp. 00\,014--00\,019.

\end{thebibliography}
